\documentclass[aps,prl,reprint,superscriptaddress]{revtex4-2}

\usepackage{jabbrv}
\usepackage{amsmath,amsfonts,amsthm,amssymb,bbm,bm,graphicx,color,enumerate,xcolor,tikz,subfigure,braket,relsize}
\usepackage{hyperref}
\usepackage{tabularray, textcomp, multirow, float,booktabs}
\usepackage[mathscr]{euscript}

\UseRawInputEncoding
\begin{document}

\title{Radio-Frequency Pseudo-Null Induced by Light in an Ion Trap}

\author{Daun Chung}
\affiliation{Dept. of Computer Science and Engineering, Seoul National University, Seoul 08826, South Korea}
\affiliation{Automation and Systems Research Institute, Seoul National University, Seoul 08826, South Korea}
\affiliation{NextQuantum, Seoul National University, Seoul 08826, South Korea}

\author{Yonghwan Cha}
\affiliation{Dept. of Computer Science and Engineering, Seoul National University, Seoul 08826, South Korea}
\affiliation{Automation and Systems Research Institute, Seoul National University, Seoul 08826, South Korea}
\affiliation{NextQuantum, Seoul National University, Seoul 08826, South Korea}

\author{Hosung Shon}
\affiliation{Dept. of Computer Science and Engineering, Seoul National University, Seoul 08826, South Korea}
\affiliation{Automation and Systems Research Institute, Seoul National University, Seoul 08826, South Korea}
\affiliation{NextQuantum, Seoul National University, Seoul 08826, South Korea}

\author{Jeonghyun Park}
\affiliation{Dept. of Computer Science and Engineering, Seoul National University, Seoul 08826, South Korea}
\affiliation{Automation and Systems Research Institute, Seoul National University, Seoul 08826, South Korea}
\affiliation{NextQuantum, Seoul National University, Seoul 08826, South Korea}

\author{Woojun Lee}
\altaffiliation{Current affiliation: Pasqal Korea, Seoul 06628, South Korea}
\affiliation{Dept. of Computer Science and Engineering, Seoul National University, Seoul 08826, South Korea}
\affiliation{Automation and Systems Research Institute, Seoul National University, Seoul 08826, South Korea}
\affiliation{Institute of Computer Technology, Seoul National University, Seoul 08826, South Korea}

\author{Kyungmin Lee}
\affiliation{Dept. of Computer Science and Engineering, Seoul National University, Seoul 08826, South Korea}
\affiliation{Automation and Systems Research Institute, Seoul National University, Seoul 08826, South Korea}
\affiliation{NextQuantum, Seoul National University, Seoul 08826, South Korea}

\author{Beomgeun Cho}
\affiliation{Dept. of Computer Science and Engineering, Seoul National University, Seoul 08826, South Korea}
\affiliation{Automation and Systems Research Institute, Seoul National University, Seoul 08826, South Korea}
\affiliation{NextQuantum, Seoul National University, Seoul 08826, South Korea}

\author{Kwangyeul Choi}
\affiliation{Dept. of Computer Science and Engineering, Seoul National University, Seoul 08826, South Korea}
\affiliation{Automation and Systems Research Institute, Seoul National University, Seoul 08826, South Korea}
\affiliation{NextQuantum, Seoul National University, Seoul 08826, South Korea}
\affiliation{Inter-University Semiconductor Research Center, Seoul National University, Seoul 08826, South Korea}

\author{Chiyoon Kim}
\affiliation{Dept. of Computer Science and Engineering, Seoul National University, Seoul 08826, South Korea}
\affiliation{Automation and Systems Research Institute, Seoul National University, Seoul 08826, South Korea}
\affiliation{NextQuantum, Seoul National University, Seoul 08826, South Korea}
\affiliation{Inter-University Semiconductor Research Center, Seoul National University, Seoul 08826, South Korea}

\author{Seungwoo Yoo}
\affiliation{Dept. of Computer Science and Engineering, Seoul National University, Seoul 08826, South Korea}
\affiliation{Automation and Systems Research Institute, Seoul National University, Seoul 08826, South Korea}
\affiliation{NextQuantum, Seoul National University, Seoul 08826, South Korea}
\affiliation{Inter-University Semiconductor Research Center, Seoul National University, Seoul 08826, South Korea}

\author{Suhan Kim}
\affiliation{Dept. of Computer Science and Engineering, Seoul National University, Seoul 08826, South Korea}
\affiliation{Automation and Systems Research Institute, Seoul National University, Seoul 08826, South Korea}
\affiliation{NextQuantum, Seoul National University, Seoul 08826, South Korea}
\affiliation{Inter-University Semiconductor Research Center, Seoul National University, Seoul 08826, South Korea}

\author{Uihwan Jeong}
\affiliation{Dept. of Computer Science and Engineering, Seoul National University, Seoul 08826, South Korea}
\affiliation{Automation and Systems Research Institute, Seoul National University, Seoul 08826, South Korea}
\affiliation{NextQuantum, Seoul National University, Seoul 08826, South Korea}
\affiliation{Inter-University Semiconductor Research Center, Seoul National University, Seoul 08826, South Korea}

\author{Jiyong Kang}
\affiliation{Dept. of Computer Science and Engineering, Seoul National University, Seoul 08826, South Korea}
\affiliation{Automation and Systems Research Institute, Seoul National University, Seoul 08826, South Korea}
\affiliation{NextQuantum, Seoul National University, Seoul 08826, South Korea}

\author{Jaehun You}
\affiliation{Dept. of Computer Science and Engineering, Seoul National University, Seoul 08826, South Korea}
\affiliation{Automation and Systems Research Institute, Seoul National University, Seoul 08826, South Korea}
\affiliation{NextQuantum, Seoul National University, Seoul 08826, South Korea}

\author{Taehyun Kim}
\thanks{taehyun@snu.ac.kr}
\affiliation{Dept. of Computer Science and Engineering, Seoul National University, Seoul 08826, South Korea}
\affiliation{Automation and Systems Research Institute, Seoul National University, Seoul 08826, South Korea}
\affiliation{NextQuantum, Seoul National University, Seoul 08826, South Korea}
\affiliation{Institute of Computer Technology, Seoul National University, Seoul 08826, South Korea}
\affiliation{Inter-University Semiconductor Research Center, Seoul National University, Seoul 08826, South Korea}
\affiliation{Institute of Applied Physics, Seoul National University, Seoul 08826, South Korea}

\begin{abstract}
In a linear radio-frequency (rf) ion trap, the rf null is the point of zero electric field in the dynamic trapping potential where the ion motion is approximately harmonic. When displaced from the rf null, the ion is superimposed by fast oscillations known as micromotion, which can be probed through motion-sensitive light-atom interactions. In this work, we report on the emergence of the rf pseudo-null, a locus of points where the ion responds to light as if it were at the true rf null, despite being displaced from it. The phenomenon is fully explained by accounting for the general two-dimensional structure of micromotion and is experimentally verified under various potential configurations, with observations in great agreement with numerical simulations. The rf pseudo-null manifests as a line in a two-dimensional parameter space, determined by the geometry of the incident light and its overlap with the motional structure of the ion. The true rf null occurs uniquely at the concurrent point of the pseudo-null lines induced by different light sources.
\end{abstract}

\maketitle

\section{Introduction}
Ion trap systems are among the leading platforms for performing quantum computation~\cite{Moses_2023, Chen_2024, Mayer_2024} and simulations~\cite{Monroe_2021, Guo_2024, Lu_2025}, owing to the stability of atomic internal states and the long coherence time of motional states within deep potentials~\cite{Auchter_2022, Spivey_2022}. In a linear radio-frequency (rf) ion trap, an oscillating quadrupole field provides confinement of motion within a two-dimensional plane through dynamic trapping~\cite{Leibfried_2003}. Here, the motion of an ion can effectively be separated into micromotion and secular motion~\cite{Berkeland_1998, Leibfried_2003}. Micromotion is the rapid motion oscillating at the frequency of the dynamic trapping field, which gives rise to a pseudopotential for secular motion at slower time scales. When the ion is at the rf null, the point of zero electric field, micromotion becomes negligible, and the secular motion can be approximated by harmonic motion. In multi-ion systems, these harmonic oscillator states can mediate ion-ion interactions, serving as the basis for numerous quantum operations~\cite{Zhu_2006, Hou_2024}.

Numerous techniques for micromotion detection and compensation have been developed~\cite{Diedrich_1987, Berkeland_1998, Ibaraki_2011, Chuah_2013, Keller_2015, Zhukas_2021, Goham_2022, Lee_2023, Hogle_2024, Arnold_2024}, primarily to minimize micromotion and position the ion at the rf null. Commonly, a simplified one-dimensional model of micromotion has been addressed, neglecting the combined structure of two-dimensional secular motion and micromotion in realistic potential configurations. A phenomenon that has remained largely unexplored due to this practice is the emergence of the rf pseudo-null, a locus of points forming a line in space, where the ion interacts with light as if there were no micromotion, even when significantly displaced from the true rf null. 

\section{Theory}
In order to describe the general two-dimensional motion of a trapped ion, we apply the invariant method used for solving time-dependent oscillator systems~\cite{Lewis_1969, Glauber_quantum, Ji_1995, Ji_1996}. Considering the case where there is an external force $\textbf{\textit{F}}(t)$ in addition to the trapping potential, the Hamiltonian of the radio-frequency trap is given as
\begin{equation}
\label{eqn_Hamiltonian_rf}
\begin{aligned}
H(t) &=
\frac{\textbf{\textit{p}}^{2}}{2M} + \frac{1}{2} M 
\textbf{\textit{x}}^\mathrm{T} \mathbf{\Omega}^{2}(t) \textbf{\textit{x}}-\textbf{\textit{F}}(t)\cdot\textbf{\textit{x}},
\end{aligned}
\end{equation}
where $\textbf{\textit{p}}$ and $\textbf{\textit{x}}$ are the momentum and position operators, and the superscript $\mathrm{T}$ indicates the transpose of a vector. The curvature of the time-dependent potential is defined through the matrix $\mathbf{\Omega}^{2}(t) = (\omega_\mathrm{rf}^{2}/4)[\mathbf{A} + 2\mathbf{Q} \mathrm{cos}(\omega_{\mathrm{rf}}t)]$~\cite{House_2008}, where $\mathbf{A}$ and $\mathbf{Q}$ are the curvature matrices generated by the static (dc) and rf potentials, respectively, whose components are given by
\begin{equation}
\label{eqn_mathieu_AQ}
\mathbf{A}_{ij} = \frac{4Ze}{M\omega_{\mathrm{rf}}^2} \left(\frac{\partial^2 \phi_\mathrm{dc}}{\partial x_{i} \partial x_{j}}\right), \ 
\mathbf{Q}_{ij} = \frac{2Ze}{M\omega_{\mathrm{rf}}^2} \left(\frac{\partial^2 \phi_\mathrm{rf}}{\partial x_{i} \partial x_{j}}\right).
\end{equation}

The constant $Z$ is the charge number ($Z = 1$ for most trapped ions), $e$ is the elementary charge, $M$ is the ion mass, and $\omega_\mathrm{rf}$ is the rf angular frequency. We assume a conventional experimental configuration where the dc and rf  potentials, $\phi_\mathrm{dc}$ and $\phi_\mathrm{rf}$, are parameterized as
\begin{equation}
\begin{gathered}
\label{eqn_mathieu_potential}
\phi_\mathrm{dc} = \frac{V_\mathrm{dc}}{2}[(\alpha - \beta)x'^{2} - (\alpha + \beta)y'^{2} + 2\beta z^{2}] \\
\phi_\mathrm{rf} = \frac{V_\mathrm{rf}}{2}\gamma (x^{2} - y^{2}).
\end{gathered}
\end{equation}

\begin{figure}[ht]
\centering
\includegraphics[width=0.7\columnwidth]{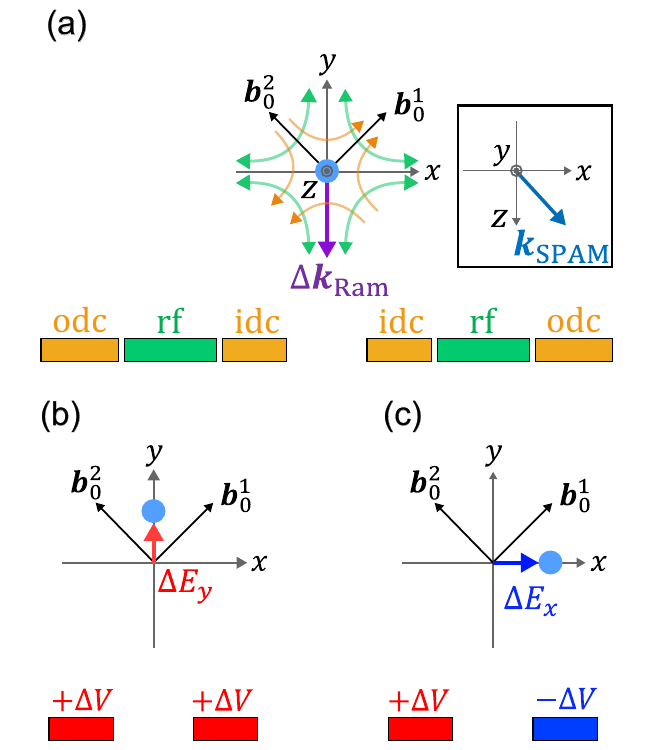}
\caption{Experimental configuration. (a) Cross-section of the surface trap. The ion, represented as a blue circle at the origin of the coordinate system, is confined by the rf (green) and dc (orange) fields generated by the electrodes of the same color. The electrode labels idc and odc stand for inner and outer dc electrodes, respectively~\cite{Chung_Choi_2024}. The principal axes of the radial modes $\textbf{\textit{b}}_{0}^{m}$ are rotated in the $xy$-plane. $\Delta\textbf{\textit{k}}_\mathrm{Ram}$ and $\textbf{\textit{k}}_\mathrm{SPAM}$ in the inset are the wave vectors of the Raman beams and the Doppler/SPAM beam. Generation of the electric fields (b) $\Delta E_{y}$, (c) $\Delta E_{x}$ by applying symmetric ($+\Delta V, +\Delta V$) and asymmetric ($+\Delta V, -\Delta V$) voltages to the idc electrodes in order to shift the ion in the $y$- and $x$-directions, respectively.}
\label{fig_config}
\end{figure}

Here, $V_\mathrm{dc}$ and $V_\mathrm{rf}$ are the dc and rf voltage amplitudes. The constraints imposed by the Laplace equation, $\nabla^2 \phi = 0$, are implicit in the curvatures $\alpha, \beta$, and $\gamma$. In general, the dc field can be rotated within the radial plane via the transformation of basis $x' = \mathrm{cos}\theta x + \mathrm{sin}\theta y, \ y' = -\mathrm{sin}\theta x + \mathrm{cos}\theta y$ (see Fig.~\ref{fig_config}). We define the two motional modes within the $xy$-plane as radial modes, and the mode confined by the static field along the $z$-direction as the axial mode. Given Eq.~(\ref{eqn_mathieu_potential}), the axial mode can be neglected as it is decoupled from the dynamic field.

The two-dimensional position operator of the radial modes in the Heisenberg picture, subject to the Hamiltonian in Eq.~(\ref{eqn_Hamiltonian_rf}), is given in Eq.~(\ref{eqn_solution_single_quantum}). The derivation of the classical solution, from which the quantum mechanical solution is constructed~\cite{Ji_1995, Ji_1996}, is presented in the Appendix. While quantization of the ion motion is not essential to elucidate the concept of the rf pseudo-null, it establishes a complete theory, enabling exact numerical simulations in the presence of both thermal and coherent motion (see Supplementary Material~\cite{supplement}).
\begin{equation}
\label{eqn_solution_single_quantum}
\begin{aligned}
\textbf{\textit{x}}_\mathrm{q}(t) = \mathlarger{\sum_{m = 1}^{2}} 
\frac{\textbf{\textit{b}}^{m}(t)}{Z_m} x_{0}^{m} \left[A_{m}^{\dag}(t) + A_{m}(t) + 2\mathrm{Re}(\alpha_{m}(t))\right]
\end{aligned}
\end{equation}

For a given mode $m$, the time-dependent vector $\textbf{\textit{b}}^{m}(t)$ is defined as $\textbf{\textit{b}}^{m}(t)=\textbf{\textit{b}}_{0}^{m} + 2 \mathrm{cos}(\omega_\mathrm{rf} t)\textbf{\textit{b}}_{1}^{m}$, where $\textbf{\textit{b}}_{0}^{m}$ is the principal axis of secular motion, and $\textbf{\textit{b}}_{1}^{m}$ is the corresponding vector of micromotion encoding both its direction and amplitude. The angular frequency $\omega_m$ is the secular frequency of mode $m$, corresponding to the harmonic component of the ion motion. We define the set of parameters $\{\omega_{m}, \textbf{\textit{b}}_{0}^{m}$, $\textbf{\textit{b}}_{1}^{m}\}$ as the secular-micromotion modes for mode $m$, and provide its formal derivation in the Appendix. The parameter $Z_m$ is a normalization constant, and $x_{0}^{m}=\sqrt{\hbar/2M\omega_{m}}$, with $\hbar$, the reduced Planck's constant. The time-varying operators $A_{m}^{\dag}(t)$ and $A_{m}(t)$ evolve as $A_{m}^{\dag}(t)= e^{i\omega_{m} t} A_{m}^{\dag}$ and $A_{m}(t)= e^{-i\omega_{m} t} A_{m}$, where $A_{m}^{\dag}$ and $A_{m}$ are the raising and lowering operators of a simple harmonic oscillator. The displacement caused by the force, $\alpha_{m}(t)$, which is the coherent component of the motion, is obtained as
\begin{equation}
\label{alpha_def}
\begin{aligned}
\alpha_{m}(t) &= e^{-i\omega_{m} t}\alpha_{m}(0) \\ &+ \frac{i}{\hbar}x_0^{m}\int_{0}^{t}dt'e^{-i\omega_{m}(t-t')}\textbf{\textit{F}}(t')\cdot \textbf{\textit{b}}_{0}^{m}
\end{aligned}
\end{equation}
subject to the initial condition, $\alpha_{m}(0) = x_{0}^{m} \textbf{\textit{F}}(0)\cdot \textbf{\textit{b}}_{0}^{m}/\hbar \omega_{m}$~\cite{Ji_1996}. In our work, we limit our analysis to the case of static forces, $\textbf{\textit{F}}(t)=\textbf{\textit{F}}(0)$, hence, $\alpha_{m}(t)=\alpha_{m}(0)$.

In order to rigorously incorporate the effects of micromotion into the light-atom interactions described by the Hamiltonian commonly used in ion trap systems~\cite{Leibfried_2003},
\begin{equation}
\label{eqn_Hamiltonian}
H'_\mathrm{LA}(t)=\frac{\hbar \Omega}{2} \left[e^{i\textbf{\textit{k}}\cdot\bm{x}(t)} e^{-i\phi(t)}\sigma^{\dag} +  \mathrm{h. c.} \right],
\end{equation}
we can simply substitute $\bm{x}(t)$ in Eq.~(\ref{eqn_Hamiltonian}) with $\textbf{\textit{x}}_\mathrm{q}(t)$ in Eq.~(\ref{eqn_solution_single_quantum}) (see Supplementary Material~\cite{supplement}). The parameters $\Omega$, $\textbf{\textit{k}}$, and $\phi(t)$ are the Rabi frequency, wave vector, and phase associated to the external radiation field, respectively. The phase can be decomposed as $\phi(t)=\delta t + \phi_{0}$ where $\delta$ is the detuning and $\phi_{0}$ is a phase offset. The operators $\sigma^{\dag}$ and $\sigma$ are the raising and lowering operators of the internal two-level system. The Hamiltonian in Eq.~(\ref{eqn_Hamiltonian}) acts on a Hilbert space spanned by $\{\ket{i} \otimes\ket{n_{1}}_{A}\otimes\ket{n_{2}}_{A} \}$ where $i \in \{0,1\}$ is the index of the two-level system, and $n_{1}, n_{2} \in \{0,1,2,...\}$ are the number-state bases of the harmonic oscillators, i.e., $A^{\dag}_{m}\ket{n_{m}}_{A}=\sqrt{n_{m}+1}\ket{n_{m}+1}_{A}$. 

\begin{figure*}[ht]
\centering
\includegraphics[width=0.9\textwidth]{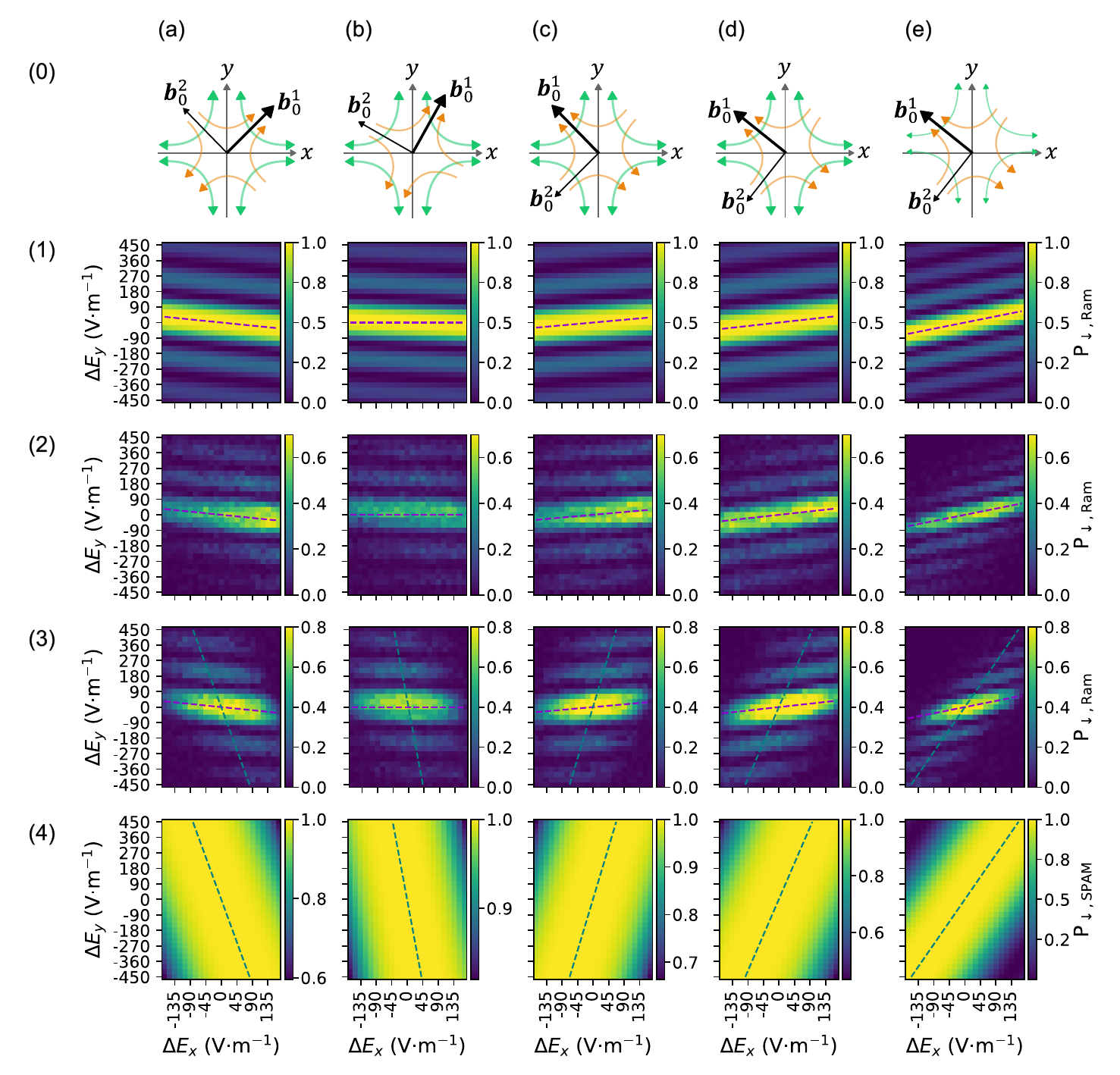}
\caption{Results of the two-dimensional displacement scan. Columns (a)-(e) group results based on different potential configurations, while rows (0)-(4) categorize data obtained under identical conditions. For each column, the secular frequencies and potential rotation angles are (a) $\{\omega_{1},\omega_{2}\}/2\pi$=\{1.79(0), 2.14(6)\} MHz, $\theta\approx\pi/4$, (b) $\{\omega_{1},\omega_{2}\}/2\pi$=\{1.87(4), 2.07(3)\} MHz, $\theta\approx\pi/3$, (c) $\{\omega_{1},\omega_{2}\}/2\pi$=\{1.82(0), 2.12(1)\} MHz, $\theta\approx3\pi/4$, (d) $\{\omega_{1},\omega_{2}\}/2\pi$=\{1.75(6), 2.22(6)\} MHz, $\theta\approx7\pi/9$, (e) $\{\omega_{1},\omega_{2}\}/2\pi$=\{1.26(8), 1.88(2)\} MHz, $\theta\approx7\pi/9$. Only the dc voltages have been varied in (a)-(d), while in (e), the rf voltage amplitude has been reduced by a factor of 0.8, keeping the dc voltages identical to those in (d). Row (0) shows the schematic depiction of potential configurations. The data in the subsequent rows are given as follows. (1) Simulations of the scan under constant Rabi drives via Raman transitions, plotting $\mathrm{P}_{\downarrow,\mathrm{Ram}}$. Experimental data of the scan with the Doppler/SPAM beam detuned by (2) -22 MHz, (3) -10 MHz from the cooling transition during Doppler cooling. Note that the color bar maxima in rows (2) and (3) are set to 0.7 and 0.8, respectively, to enable a comparison of the relative Raman transition strengths for each case. (4) Simulation of the scan conducted with the Doppler/SPAM beam, plotting $\mathrm{P}_{\downarrow,\mathrm{SPAM}}$. The theoretical rf pseudo-null lines induced by the Raman beams and Doppler/SPAM beam are represented as dashed lines in purple and teal, respectively.}
\label{fig_2Dscan}
\end{figure*}

The rf pseudo-null arises from the spatial phase term
\begin{equation}
\label{phi_mm}
\begin{aligned}
\phi_\mathrm{mm}(t) = 2{\sum_{m = 1}^{2}}(\textbf{\textit{k}}\cdot\textbf{\textit{b}}^{m}(t)/{Z_m}) x_{0}^{m}\mathrm{Re}(\alpha_{m}(0)),
\end{aligned}
\end{equation}
which induces phase modulation in the light-atom interaction (see the Appendix). Attributed to contribution from the two modes, the net phase can cancel out, $\phi_\mathrm{mm}(t)=0$, with the condition determined by the orientation of the wave vector $\textbf{\textit{k}}$ and its overlap with the ion's motional structure. Effectively, micromotion is hidden from the incident light, and the ion behaves as if it were a simple harmonic oscillator, even when superimposed by large micromotion. The cancellation of the spatial phase components is coherent, as $\phi_\mathrm{mm}(t)$ is independent of the operators $A_{m}^{\dag}$ and $A_{m}$, leading to distinct signatures even when the ion is highly thermal, as demonstrated later.

\section{Experimental Setup}
The rf pseudo-null is visualized through a series of experiments extended from the dc scanning method originally developed for the compensation of micromotion along the direction normal to the chip surface~\cite{Lee_2023}. Dubbed the two-dimensional displacement scan, a single experimental sequence is executed in the following order. First, dc voltages are set to position the ion at the true rf null. Then, the ion is displaced from the rf null within the $xy$-plane by applying incremental voltages to the inner dc (idc) electrodes, as shown in Fig.~\ref{fig_config}. This amounts to applying a constant force to the ion through an external electric field, $\textbf{\textit{F}}(t)=e\Delta \textbf{\textit{E}}_{0}=e(\Delta E_{x}\bar{\textbf{\textit{x}}} + \Delta E_{y}\bar{\textbf{\textit{y}}})$. At the new equilibrium position, the ion is Doppler-cooled and the internal state is initialized to state $\ket{0}=\ket{\uparrow}$. At each displacement, a motion-sensitive Rabi flopping is driven for a duration of $T=\pi/\Omega$, tuned at the carrier frequency, $\delta=0$, after which the probability of the state $\ket{1}=\ket{\downarrow}$, $\mathrm{P}_{\downarrow}$, is measured~\cite{Lee_2023}. Away from the true rf null, micromotion increases and modulates the phase of light seen by the ion, given by $\phi_\mathrm{mm}(t)$ in Eq.~(\ref{phi_mm}), effectively modifying the Rabi frequency (see the Appendix). The pattern of $\mathrm{P}_{\downarrow}$ extracted when the displacement of the ion is scanned over a two-dimensional region is directly dependent on $\phi_\mathrm{mm}(t)$, with the orientation of light, $\textbf{\textit{k}}$, most significantly altering the pattern. The result is independent of the Rabi frequency as long as it is much smaller than the secular frequencies, $\Omega \ll \omega_{1}, \omega_{2}$.

The experiments are performed with a silicon-based microfabricated surface trap~\cite{Chung_Choi_2024}. Hyperfine levels in the S-orbital of the $^{171}$Yb$^+$ ion, $\ket{\uparrow}=\ket{^{2}S_{1/2}, F=0,  m_{F}=0}$ and $\ket{\downarrow}=\ket{^{2}S_{1/2}, F=1, m_{F}=0}$, are used as the internal states that undergo Raman transitions via two counter-propagating 355-nm beams derived from a mode-locked laser~\cite{Olmschenk_2007, Mizrahi_2014}. In this case, the wave vector $\textbf{\textit{k}}$ in Eq.~(\ref{eqn_Hamiltonian}) is replaced with $\Delta\textbf{\textit{k}}_{\mathrm{Ram}}$, the difference of the wave vectors from the two Raman beams, as shown in Fig.~\ref{fig_config} (a). The effective wave vector $\Delta\textbf{\textit{k}}_{\mathrm{Ram}}=-2\times(2\pi/\lambda_{355})\bar{\textbf{\textit{y}}}$ does not overlap with the $z$-axis, making the Raman transition sensitive only to the radial motion of the ion. A 370-nm laser, referred to as the Doppler/SPAM beam, is used for Doppler cooling as well as state preparation and measurement (SPAM)~\cite{Olmschenk_2007}. Its orientation is given by the wave vector $\textbf{\textit{k}}_{\mathrm{SPAM}}=(2\pi/\lambda_{370})(\bar{\textbf{\textit{x}}}+\bar{\textbf{\textit{z}}})/\sqrt{2}$, as shown in the inset of Fig.~\ref{fig_config} (a). The main cooling transition occurs between the state $\ket{\downarrow}$ and a state $\ket{e}=\ket{^{2}P_{1/2}, F=0, m_{F}=0}$ in the P-orbital~\cite{Olmschenk_2007}.

\section{Results}
The results of the two-dimensional displacement scan are plotted in Fig.~\ref{fig_2Dscan}. The columns (a)-(e) display data obtained from different potential configurations, which are represented by the diagrams in row (0). The parameter values of $\alpha$, $\beta$, and $\gamma$ in Eq.~(\ref{eqn_mathieu_potential}) for each condition are provided in the Supplemental Material~\cite{supplement}. In the captions of Fig.~\ref{fig_2Dscan}, we specify the secular frequency of each mode, and the potential rotation angle $\theta$, defined as the angle between the $x$-axis and the principal axis $\textbf{\textit{b}}_{0}^{1}$. Mode $m=1$ is designated as the mode with the smaller secular frequency, with its axis $\textbf{\textit{b}}_{0}^{1}$ emphasized in thicker lines in the diagrams. The numerical values of the secular-micromotion modes are also listed in the Supplemental Material~\cite{supplement}. Throughout Fig.~\ref{fig_2Dscan} (a)-(d), the dc potential parameters are varied while the rf potential parameters are kept constant. In (e), on the other hand, the dc potential parameters are identical to those in (d), while the rf voltage amplitude is reduced. In all cases, $\omega_\mathrm{rf}=2\pi \times 22.14 \  \mathrm{MHz}$.

Row (1) of Fig.~\ref{fig_2Dscan} exhibits simulation data of the two-dimensional scan. At each displacement set by the electric field $\textbf{\textit{E}}_{0}=\Delta E_{x}\bar{\textbf{\textit{x}}}+ \Delta E_{y}\bar{\textbf{\textit{y}}}$, the ion evolves from an initial state $\ket{\psi(t=0)}=\ket{\uparrow}\otimes\ket{0_{1}}_{A}\otimes\ket{0_{2}}_{A}$. Distinct global patterns, dependent on the experimental geometry, are predicted to occur and are clearly observed experimentally, as displayed in row (2) of Fig.~\ref{fig_2Dscan}. A key feature is the emergence of the rf pseudo-null line, a continuous set of points where $\phi_\mathrm{mm}(t)=0$, indicated by purple dashed lines. The maximal values of the excitation probability, $\mathrm{P}_{\downarrow,\mathrm{Ram}}$, are distributed along this line. Note that we indicate $\mathrm{P}_{\downarrow}$ occurring from the Raman transition as $\mathrm{P}_{\downarrow,\mathrm{Ram}}$ in order to distinguish it from the excitation probability associated with the Doppler/SPAM beam, $\mathrm{P}_{\downarrow,\mathrm{SPAM}}$, introduced later. A Bessel-type pattern occurs in $\mathrm{P}_{\downarrow,\mathrm{Ram}}$ with respect to this pseudo-null line, even when the ion motion is highly thermal, due to the nature of coherent sinusoidal phase modulation~\cite{Gaebler_2021,Lee_2023,Lysne_2024}.

The experimental data in rows (2) and (3) of Fig.~\ref{fig_2Dscan} are acquired under identical conditions, but with the Doppler/SPAM beam detuned by $-22$ MHz and $-10$ MHz from the cooling transition during Doppler cooling, respectively. The former enables a relatively uniform cooling condition throughout the region since the detuning is set to the rf frequency, efficiently cooling the ion via micromotion sidebands~\cite{Cirac_1994, Berkeland_1998}. The latter, on the other hand, corresponds to the optimal Doppler cooling condition~\cite{Cirac_1994,Berkeland_1998,Eschner_2003}, given the linewidth of the cooling transition of approximately 20 MHz. Here, the ion is cooled down to states with smaller mean phonon numbers, leading to higher values of $\mathrm{P}_{\downarrow,\mathrm{Ram}}$ around the true rf null compared to the former, due to less thermal motional decoherence~\cite{Schulz_2008}. However, we observe that the signals become localized in a manner that is not fully accounted for by the simulation results in row (1) of Fig.~\ref{fig_2Dscan}. 

\begin{figure}[h]
\centering
\includegraphics[width=1\columnwidth]{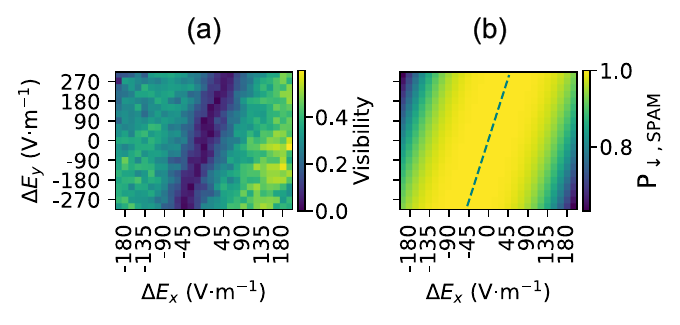}
\caption{Two-dimensional displacement scan conducted with the Doppler/SPAM beam. (a) Visibility extracted from photon correlation experiments. (b) Simulation results of the Hamiltonian in Eq.~(\ref{eqn_Hamiltonian}) acting on the two-level system $\ket{0}=\ket{\downarrow}$, $\ket{1}=\ket{e}$, and $\textbf{\textit{k}}$ replaced by $\textbf{\textit{k}}_{\mathrm{SPAM}}$.}
\label{fig_correlation}
\end{figure}

This secondary feature actually originates from a different rf pseudo-null line induced by the Doppler/SPAM beam, given by the geometry in Fig.~\ref{fig_config} (a).  With the two-level states defined as $\ket{0}=\ket{\downarrow}$, $\ket{1}=\ket{e}$, and the wave vector replaced with that of the 370-nm laser, $\textbf{\textit{k}}_{\mathrm{SPAM}}$, Eq.~(\ref{eqn_Hamiltonian}) describes the system Hamiltonian involved in a dissipative scattering process~\cite{DeVoe_1989,Dum_1992,Gardiner_2015}. Neglecting scattering, we simulate the two-dimensional displacement scan with respect to the Doppler/SPAM beam, whose results are shown in row (4) of Fig.~\ref{fig_2Dscan}. Here, $\mathrm{P}_{\downarrow,\mathrm{SPAM}}$ is equivalent to the excitation probability of the state $\ket{e}$, $\mathrm{P}_{e}$, which is a measure of fluorescence intensity. The subscript $\downarrow$ is used to reflect the experimental aspect that the standard state detection scheme distinguishes the state $\ket{\downarrow}$ from $\ket{\uparrow}$ by collecting fluorescence emitted during near-resonant driving of the cooling transition~\cite{Olmschenk_2007}. Note that the quantity $\epsilon=1-\mathrm{P}_{\downarrow,\mathrm{SPAM}}$ does not correspond to the detection probability of the state $\ket{\uparrow}$, but rather the detection error associated with the loss of fluorescence due to altered transition strengths between the states $\ket{\downarrow}$ and $\ket{e}$ by micromotion. In the Supplementary Material~\cite{supplement}, we discuss the general situation where scattering is present, in contrast to simulations involving only the system Hamiltonian, and present experimental results of the two-dimensional displacement scans of $\mathrm{P}_{\downarrow,\mathrm{SPAM}}$ conducted identically to that of $\mathrm{P}_{\downarrow,\mathrm{Ram}}$. In Fig.~\ref{fig_2Dscan} row (4), the rf pseudo-null induced by the Doppler/SPAM beam is indicated by dashed lines colored in teal. In the region with high $\mathrm{P}_{\downarrow,\mathrm{SPAM}}$ values, the ion scatters more photons, resulting in  more effective Doppler cooling to lower temperatures~\cite{Javanainen_1980}, and higher detection fidelity~\cite{Olmschenk_2007}.

There is a more direct way to observe the pseudo-null line induced by the Doppler/SPAM beam through a commonly used micromotion compensation technique, typically termed photon correlation~\cite{Diedrich_1987, Keller_2015}. We conduct a two-dimensional displacement scan, where the visibility from a photon correlation experiment is extracted and plotted at each point. The visibility, defined as $(I_{\mathrm{max}}-I_{\mathrm{min}})/(I_{\mathrm{max}}+I_{\mathrm{min}})$, with $I_{\mathrm{max}}$ and $I_{\mathrm{min}}$ denoting the maximum and minimum photon counts recorded over one rf oscillation period, serves as an indicator of micromotion seen by the laser. The experimental result, obtained under the potential configuration in Fig.~\ref{fig_2Dscan} (c), is presented in Fig.~\ref{fig_correlation} (a). Here, we clearly see the rf pseudo-null line as the locus of data points where the visibility reaches zero. This is in agreement with simulation data, shown in Fig.~\ref{fig_correlation} (b), where the rf pseudo-null line induced by the Doppler/SPAM beam is indicated by a dashed line. The pseudo-null line appears sharper in (a) because, while the photon correlation method directly probes the strength of micromotion, the Hamiltonian simulation results in (b) show how the excitation probability is modified in the presence of micromotion.

It follows that the data in row (3) of Fig.~\ref{fig_2Dscan} visualizes the combined pattern of the two rf pseudo-null lines induced by both the Raman and Doppler/SPAM beams, with the influence of $\mathrm{P}_{\downarrow,\mathrm{SPAM}}$ implicitly reflected in $\mathrm{P}_{\downarrow,\mathrm{Ram}}$ through the SPAM fidelity, which manifests as a gradual fading of signals at the edges of the high-$\mathrm{P}_{\downarrow,\mathrm{SPAM}}$ region. The true rf null lies at the common intersection of the pseudo-null lines induced by the different beams. This concurrent point, tuned to $\Delta E_{x}=\Delta E_{y}=0$, is indeed where the highest excitation probability $\mathrm{P}_{\downarrow,\mathrm{Ram}}$ occurs empirically, under the optimal Doppler cooling condition. Our results imply that performing the two-dimensional displacement scan can uniquely identify the true rf null, provided that at least two pseudo-null lines are observed. These pseudo-null lines do not necessarily have to be induced by mutually orthogonal beams, as the localization of signals occurs at the concurrent point as long as they are non-collinear. To our knowledge, this is the only self-contained method that achieves complete determination of the true rf null in the radial plane using a single ion.

\section{Conclusion}
To summarize, we introduce the notion of the rf pseudo-null, a phenomenon overlooked in ion trap systems due to an insufficient understanding of the general two-dimensional ion motion beyond the pseudopotential regime. Our theory accurately predicts the emergence of the rf pseudo-null as a consequence of coherent light-atom interactions, consistent with observations from the two-dimensional displacement scan experiments. This approach also serves as an experimentally feasible method for locating the true rf null at the common intersection of multiple pseudo-null lines, requiring at minimum a single motion-sensitive beam in addition to the primary beam used for mandatory SPAM operations.

\section{Acknowledgment}
This work has been supported by the Institute for Information \& Communications Technology Planning \& Evaluation (IITP) grant (No. 2022-0-01040), and the National Research Foundation of Korea (NRF) grant (No. RS-2020-NR049232, No. RS-2024-00442855, No. RS-2024-00413957), all of which are funded by the Korean government (MSIT).

\appendix
\section*{Appendix}

\section{Secular-Micromotion Modes}
\label{sec_mode_structure}
Let us decompose the Hamiltonian in Eq.~(\ref{eqn_Hamiltonian_rf}) as follows.
\begin{equation}
\label{eqn_hamiltonian_rf_de}
\begin{gathered}
H(t)=H_{0}(t)+V(t) \\
H_{0}(t) = \frac{\textbf{\textit{p}}^{2}}{2M} + \frac{1}{2} M 
\textbf{\textit{x}}^\mathrm{T} \mathbf{\Omega}^{2}(t) \textbf{\textit{x}} \\
V(t) = -\textbf{\textit{F}}(t)\cdot\textbf{\textit{x}}
\end{gathered}
\end{equation}

The secular-micromotion mode structure can be derived by solving the classical equations of motion subject to the Hamiltonian $H_{0}(t)$, which simply results in the Mathieu equation~\cite{Leibfried_2003}:
\begin{equation}
\label{eqn_mathieu_single}
\begin{gathered}
\frac{d^{2}\textbf{\textit{x}}}{dt^{2}} + \mathbf{\Omega}^{2}(t)\textbf{\textit{x}} = \textbf{0}.
\end{gathered}
\end{equation}

Assuming an ansatz

\begin{equation}
\label{eqn_ansatz_single}
\textbf{\textit{f}}(t) = e^{i \frac{\lambda}{2} \omega_\mathrm{rf} t} \mathlarger{\sum_{n=-\infty}^{\infty}} e^{i n \omega_\mathrm{rf} t} \textbf{\textit{b}}_{n},
\end{equation}
we obtain the relation \cite{House_2008}
\begin{equation}
\label{eqn_relation_single}
-\mathbf{D}_{n}\textbf{\textit{b}}_{n} + \textbf{\textit{b}}_{n - 1} + \textbf{\textit{b}}_{n + 1} = \textbf{0},
\end{equation}
where $\mathbf{D}_{n} = \mathbf{Q}^{-1}[(2n + \lambda)^{2}\mathbf{I} -\mathbf{A}]$, with $\mathbf{I}$ being the identity matrix. Stable solutions occur under the condition $|\textbf{\textit{b}}_{0}| \gg |\textbf{\textit{b}}_{\pm 1}| \gg |\textbf{\textit{b}}_{\pm 2}| \approx 0$. This results in the following set of equations
\begin{equation}
\begin{gathered}
\label{eqn_reduced}
-\mathbf{D}_{0}\textbf{\textit{b}}_{0} + \textbf{\textit{b}}_{-1} + \textbf{\textit{b}}_{+1} = \textbf{0} \\
\textbf{\textit{b}}_{\pm 1} = \mathbf{D}_{\pm 1}^{-1}\textbf{\textit{b}}_{0},
\end{gathered}
\end{equation}
which can be expressed as a single equation as
\begin{equation}
\label{eqn_single}
[\mathbf{D}_{0} - \mathbf{D}_{-1}^{-1} - \mathbf{D}_{+1}^{-1}]\textbf{\textit{b}}_{0} = \textbf{0}.
\end{equation}

It can be seen that Eq.~(\ref{eqn_single}) not only determines the principal axes of the radial modes, $\textbf{\textit{b}}_{0}$, but also $\textbf{\textit{b}}_{\pm 1}$ through Eq.~(\ref{eqn_reduced}), which we define as micromotion modes. In the lowest order approximation, $\lambda \ll |2|$, we define $\mathbf{D}_{1} \equiv \mathbf{D}_{\pm 1} \approx \mathbf{Q}^{-1}(4\mathbf{I} -\mathbf{A})$, so that Eq.~(\ref{eqn_single}) is reduced to an eigenvalue problem with respect to $\lambda$,
\begin{equation}
\label{eqn_eigenvalue}
[\mathbf{D}_{0} - 2\mathbf{D}_{1}^{-1}]\textbf{\textit{b}}_{0} = \textbf{0},
\end{equation}
which results in the diagonalization of the matrix $\mathbf{G} = \mathbf{A} + 2 \mathbf{Q} \mathbf{D}_{1}^{-1}$. Then, the solution can be obtained by taking the real part of
\begin{equation}
\label{eqn_solution_single}
\begin{aligned}
\textbf{\textit{f}}(t) &= \mathlarger{\sum_{m=1}^{2}} \textbf{\textit{b}}^{m}(t) u_m e^{i \omega_{m} t}, \\ 
\textbf{\textit{b}}^{m}(t) &=\textbf{\textit{b}}_{0}^{m} + 2 \mathrm{cos}(\omega_\mathrm{rf} t) \textbf{\textit{b}}_{1}^{m} \\ 
&= \left(I + \frac{q}{2} \mathrm{cos}(\omega_\mathrm{rf} t) \mathbf{D}^{-1} \right) \textbf{\textit{b}}_{0}^{m}
\end{aligned}
\end{equation}
where $u_{m}$ is the amplitude of mode $m$, $\omega_{m} = \lambda_{m} \omega_\mathrm{rf} / 2$ and $\textbf{\textit{b}}_{0}^{m}$ are the secular frequencies and corresponding principal axes of mode $m$, respectively, and $\mathbf{D} \equiv (q/4) \mathbf{D}_{1}$, with $\textbf{\textit{b}}_{1}^{m}=\mathbf{D}_{1}^{-1}\textbf{\textit{b}}_{0}^{m}$. Note that $q = 2eV_\mathrm{rf}\gamma/M\omega_{\mathrm{rf}}^2$ from Eqs.~(\ref{eqn_mathieu_AQ}) and~(\ref{eqn_mathieu_potential}). In realistic experimental configurations where the dc potential is rotated, the off-diagonal terms of the matrix $\mathbf{D}_{1}$ are non-zero, so that $\textbf{\textit{b}}_{1}^{m}$ is non-collinear to $\textbf{\textit{b}}_{0}^{m}$, which is a feature that is commonly overlooked in many studies on micromotion. Although it is straightforward to extend the formalism to include higher order micromotion modes $\textbf{\textit{b}}^{m}_{\pm 2}, \textbf{\textit{b}}^{m}_{\pm 3},...$, their magnitude is computed to be negligible for typically occurring values of $q$ within the range $0.1<q<0.3$.

\section{The Forced Time-dependent Oscillator}
\label{sec_forced}
The particular solution to the classical equations of motion for mode $m$ when $V(t)$ in Eq.~(\ref{eqn_hamiltonian_rf_de}) is present,
\begin{equation}
\label{eqn_mathieu_single_forced}
\begin{gathered}
\frac{d^{2}\textbf{\textit{x}}}{dt^{2}} + \mathbf{\Omega}^{2}(t)\textbf{\textit{x}} = \frac{\textbf{\textit{F}}(t)}{M},
\end{gathered}
\end{equation}
can be obtained by assuming an ansatz $\textbf{\textit{x}}_{\mathrm{p}}(t)=u_{0}^{m}(t)\textbf{\textit{b}}^{m}(t)$, which leads to the following equation for $u_{0}^{m}(t)$, 
\begin{equation}
\label{eqn_mathieu_single_forced_particular}
\begin{gathered}
\frac{d^{2}u_{0}^{m}}{dt^{2}} + \mathbf{\Omega}^{2}(t)u_{0}^{m} = \frac{\textbf{\textit{F}}(t) \cdot \textbf{\textit{b}}_{0}^{m}}{M},
\end{gathered}
\end{equation}
under the lowest order approximation for stable solutions, $\mathrm{cos(\omega_\mathrm{rf}t)}=\mathrm{sin(\omega_\mathrm{rf}t)}=0, \ \mathrm{cos^{2}(\omega_\mathrm{rf}t)}=1/2$. The solution to Eq.~(\ref{eqn_mathieu_single_forced_particular}) can be expressed via the Green function method as 
\begin{equation}
\label{sol_mathieu_single_forced_particular}
\begin{aligned}
u_\mathrm{0}^{m}(t) &= u_\mathrm{0}^{m}(0) \mathrm{cos}(\omega_{m} t) \\ &+ \frac{1}{M \omega_{m}}\int_{0}^{t}dt'\mathrm{sin}\left(\omega_{m}(t-t')\right)\textbf{\textit{F}}(t')\cdot \textbf{\textit{b}}_{0}^{m},
\end{aligned}
\end{equation}
with the initial condition, $u_\mathrm{0}^{m}(0) = \textbf{\textit{F}}(0)\cdot \textbf{\textit{b}}_{0}^{m} / M \omega_{m}^{2}$. The initial condition is chosen so that for a constant force, $\textbf{\textit{F}}(t)=\textbf{\textit{F}}_0$, we get $u_\mathrm{0}^{m}(t)=u_\mathrm{0}^{m}(0)$.  Then, using Eqs.~(\ref{eqn_solution_single}) and (\ref{sol_mathieu_single_forced_particular}), the total classical solution can be expressed as~\cite{Berkeland_1998}
\begin{equation}
\label{eqn_solution_single_c}
\begin{aligned}
\textbf{\textit{x}}_{\mathrm{c}}(t) &= \mathlarger{\sum_{m=1}^{2}} \textbf{\textit{b}}^{m}(t) \left[u_\mathrm{1}^{m} \mathrm{cos}(\omega_{m} t) + u_\mathrm{0}^{m}(t)\right].
\end{aligned}
\end{equation}

\section{The Spatial Phase}
Quantization of the solution $\textbf{\textit{x}}_{\mathrm{c}}(t)$ in Eq.~(\ref{eqn_solution_single_c}) leads to the operator $\textbf{\textit{x}}_{\mathrm{q}}(t)$ in Eq.~(\ref{eqn_solution_single_quantum}) (see the Supplementary Material~\cite{supplement}). The spatial phase that arises in the light-atom interaction described by Eq.~(\ref{eqn_Hamiltonian}) is decomposed as
\begin{equation}
\label{eqn_kx}
\begin{gathered}
\textbf{\textit{k}}\cdot\bm{x}_\mathrm{q}(t)= \phi_\mathrm{sm}(t) + \phi_\mathrm{mm}(t)
\end{gathered}
\end{equation}
where $\phi_\mathrm{sm}(t)$ is the harmonic oscillator term arising from secular motion
\begin{equation}
\label{eqn_phase_secular}
\begin{gathered}
\\ \phi_\mathrm{sm}(t)=\mathlarger{\sum_{m = 1}^{2}} 
\frac{\textbf{\textit{k}}\cdot\textbf{\textit{b}}^{m}_{0}}{Z_m} x_{0}^{m} \left(A_{m}^{\dag}e^{i\omega_{m}t} + A_{m}e^{-i\omega_{m}t}\right),
\end{gathered}
\end{equation}
and $\phi_\mathrm{mm}$(t), originating from micromotion, can be further reduced to the intrinsic and excess micromotion components
\begin{equation}
\label{eqn_phase_micromotion}
\begin{gathered}
\phi_\mathrm{mm}(t) = \phi_\mathrm{in}(t) + \phi_\mathrm{ex}(t),
\\ \phi_\mathrm{in}(t) = 2\mathrm{cos}(\omega_\mathrm{rf}t){\sum_{m = 1}^{2}} 
\frac{\textbf{\textit{k}}\cdot\textbf{\textit{b}}^{m}_{1}}{Z_m} x_{0}^{m}\left(A_{m}^{\dag}e^{i\omega_{m}t} + A_{m}e^{-i\omega_{m}t}\right)
\\ \phi_\mathrm{ex}(t) = 2{\sum_{m = 1}^{2}} 
\frac{\textbf{\textit{k}}\cdot\textbf{\textit{b}}^{m}(t)}{Z_m} x_{0}^{m}\mathrm{Re}(\alpha_{m}(t)).
\end{gathered}
\end{equation}
In numerical simulations, the intrinsic part $\phi_\mathrm{in}(t)$ is neglected, $\phi_\mathrm{in}(t)\approx 0$, because it unnecessarily complicates the computation despite its small significance. Then, since the excess component $\phi_\mathrm{ex}(t)$ is a constant, the effect of micromotion is to induce a coherent phase modulation to the light-atom interaction through the factor $\mathrm{exp}(i\phi_\mathrm{mm}(t))\approx\mathrm{exp}(i\phi_\mathrm{ex}(t))$, which is the physical basis of displacement-dependent Rabi frequency variations.

\clearpage
\section*{Supplementary Material}
\title{Supporting Online Material for \\ Radio-Frequency Pseudo-Null Induced by Light in an Ion Trap}

\author{Daun Chung}
\affiliation{Dept. of Computer Science and Engineering, Seoul National University, Seoul 08826, South Korea}
\affiliation{Automation and Systems Research Institute, Seoul National University, Seoul 08826, South Korea}
\affiliation{NextQuantum, Seoul National University, Seoul 08826, South Korea}

\author{Yonghwan Cha}
\affiliation{Dept. of Computer Science and Engineering, Seoul National University, Seoul 08826, South Korea}
\affiliation{Automation and Systems Research Institute, Seoul National University, Seoul 08826, South Korea}
\affiliation{NextQuantum, Seoul National University, Seoul 08826, South Korea}

\author{Hosung Shon}
\affiliation{Dept. of Computer Science and Engineering, Seoul National University, Seoul 08826, South Korea}
\affiliation{Automation and Systems Research Institute, Seoul National University, Seoul 08826, South Korea}
\affiliation{NextQuantum, Seoul National University, Seoul 08826, South Korea}

\author{Jeonghyun Park}
\affiliation{Dept. of Computer Science and Engineering, Seoul National University, Seoul 08826, South Korea}
\affiliation{Automation and Systems Research Institute, Seoul National University, Seoul 08826, South Korea}
\affiliation{NextQuantum, Seoul National University, Seoul 08826, South Korea}

\author{Woojun Lee}
\altaffiliation{Current affiliation: Pasqal Korea, Seoul 06628, South Korea}
\affiliation{Dept. of Computer Science and Engineering, Seoul National University, Seoul 08826, South Korea}
\affiliation{Automation and Systems Research Institute, Seoul National University, Seoul 08826, South Korea}
\affiliation{Institute of Computer Technology, Seoul National University, Seoul 08826, South Korea}

\author{Kyungmin Lee}
\affiliation{Dept. of Computer Science and Engineering, Seoul National University, Seoul 08826, South Korea}
\affiliation{Automation and Systems Research Institute, Seoul National University, Seoul 08826, South Korea}
\affiliation{NextQuantum, Seoul National University, Seoul 08826, South Korea}

\author{Beomgeun Cho}
\affiliation{Dept. of Computer Science and Engineering, Seoul National University, Seoul 08826, South Korea}
\affiliation{Automation and Systems Research Institute, Seoul National University, Seoul 08826, South Korea}
\affiliation{NextQuantum, Seoul National University, Seoul 08826, South Korea}

\author{Kwangyeul Choi}
\affiliation{Dept. of Computer Science and Engineering, Seoul National University, Seoul 08826, South Korea}
\affiliation{Automation and Systems Research Institute, Seoul National University, Seoul 08826, South Korea}
\affiliation{Inter-University Semiconductor Research Center, Seoul National University, Seoul 08826, South Korea}
\affiliation{NextQuantum, Seoul National University, Seoul 08826, South Korea}

\author{Chiyoon Kim}
\affiliation{Dept. of Computer Science and Engineering, Seoul National University, Seoul 08826, South Korea}
\affiliation{Automation and Systems Research Institute, Seoul National University, Seoul 08826, South Korea}
\affiliation{Inter-University Semiconductor Research Center, Seoul National University, Seoul 08826, South Korea}
\affiliation{NextQuantum, Seoul National University, Seoul 08826, South Korea}

\author{Seungwoo Yoo}
\affiliation{Dept. of Computer Science and Engineering, Seoul National University, Seoul 08826, South Korea}
\affiliation{Automation and Systems Research Institute, Seoul National University, Seoul 08826, South Korea}
\affiliation{Inter-University Semiconductor Research Center, Seoul National University, Seoul 08826, South Korea}
\affiliation{NextQuantum, Seoul National University, Seoul 08826, South Korea}

\author{Suhan Kim}
\affiliation{Dept. of Computer Science and Engineering, Seoul National University, Seoul 08826, South Korea}
\affiliation{Automation and Systems Research Institute, Seoul National University, Seoul 08826, South Korea}
\affiliation{Inter-University Semiconductor Research Center, Seoul National University, Seoul 08826, South Korea}
\affiliation{NextQuantum, Seoul National University, Seoul 08826, South Korea}

\author{Uihwan Jeong}
\affiliation{Dept. of Computer Science and Engineering, Seoul National University, Seoul 08826, South Korea}
\affiliation{Automation and Systems Research Institute, Seoul National University, Seoul 08826, South Korea}
\affiliation{Inter-University Semiconductor Research Center, Seoul National University, Seoul 08826, South Korea}
\affiliation{NextQuantum, Seoul National University, Seoul 08826, South Korea}

\author{Jiyong Kang}
\affiliation{Dept. of Computer Science and Engineering, Seoul National University, Seoul 08826, South Korea}
\affiliation{Automation and Systems Research Institute, Seoul National University, Seoul 08826, South Korea}
\affiliation{NextQuantum, Seoul National University, Seoul 08826, South Korea}

\author{Jaehun You}
\affiliation{Dept. of Computer Science and Engineering, Seoul National University, Seoul 08826, South Korea}
\affiliation{Automation and Systems Research Institute, Seoul National University, Seoul 08826, South Korea}
\affiliation{NextQuantum, Seoul National University, Seoul 08826, South Korea}

\author{Taehyun Kim}
\thanks{taehyun@snu.ac.kr}
\affiliation{Dept. of Computer Science and Engineering, Seoul National University, Seoul 08826, South Korea}
\affiliation{Automation and Systems Research Institute, Seoul National University, Seoul 08826, South Korea}
\affiliation{Institute of Computer Technology, Seoul National University, Seoul 08826, South Korea}
\affiliation{Inter-University Semiconductor Research Center, Seoul National University, Seoul 08826, South Korea}
\affiliation{Institute of Applied Physics, Seoul National University, Seoul 08826, South Korea}
\affiliation{NextQuantum, Seoul National University, Seoul}

\maketitle

\section{The invariant method: Quantization of secular-micromotion modes}
\label{sec_invariant}

We first review the invariant method applied to the quantum mechanical description of an undriven ion trapped along one dimension~\cite{Ji_1995}, in which case the Hamiltonian is reduced to
\begin{equation}
\label{eqn_Hamiltonian_sup_rf_1d}
\begin{gathered}
H_{0}(t) = \frac{p_{x}^{2}}{2M} + \frac{1}{2} M \Omega_{x}^{2}(t) x^{2}, \\
\Omega_{x}^{2}(t) =\frac{\omega_\mathrm{rf}^{2}}{4}\left(a_x + 2q_x \mathrm{cos}(\omega_\mathrm{rf} t)\right)
\end{gathered}
\end{equation}

The position operator in the Heisenberg picture is obtained as~\cite{Lee_Chung_2024}
\begin{equation}
\label{eqn_pos_1d}
x(t)=\sqrt{\frac{\hbar g_{-}(t)}{2\omega_{x}}}\left(e^{i\omega(t)}a_{x}^{\dag}+e^{-i\omega(t)}a_{x}\right)
\end{equation}
where $\omega_x$ is the invariant of the system and $g_{-}(t)=|f(t)|^{2}/M$ is obtained from the classical solution of the one dimensional system $f(t)$ from Eq.~(15) in the Appendix, subject to the initial condition $f(0)=1$ and $\dot{f}(0)=i\omega_x$~\cite{Glauber_quantum}:
\begin{equation}
\label{eqn_f_sol}
f(t)= e^{i\omega_x t}\frac{1+\frac{q_x}{2}\mathrm{cos}(\omega_\mathrm{rf}t)}{1+\frac{q_x}{2}}.
\end{equation}

The term $\omega(t)$, obtained as
\begin{equation}
\label{eqn_omega_x}
\omega(t)=\omega_x {\mathlarger{\int}_{0}^{t}} dt' \frac{1}{Mg_{-}(t^{'})},
\end{equation}
is reduced to $\omega(t)=\omega_{x}t$ when the terms resulting in factors of $q_{x}^2$ are approximated to 1, i.e., $|f(t)|^2 \approx 1$. In this limit, the expression in Eq.~(\ref{eqn_pos_1d}) becomes 
\begin{equation}
\label{eqn_pos_1d_new}
x(t)= \left(\frac{1+\frac{q_x}{2}\mathrm{cos}(\omega_\mathrm{rf}t)}{1+\frac{q_x}{2}}\right) 
x_{0} \left(e^{i\omega_{x}t}a_{x}^{\dag}+e^{-i\omega_{x}t}a_{x}\right),
\end{equation}
where $x_{0}=\sqrt{\hbar/2M\omega_{x}}$. This corresponds to the classical solution upon taking the expectation value 
\begin{equation}
\label{eqn_pos_1d_exp}
\bra{u_{x}}x(t)\ket{u_{x}} \propto \left(1 + \frac{q_{x}}{2} \mathrm{cos}(\omega_\mathrm{rf} t)\right) \mathrm{cos}(\omega_{x} t)
\end{equation}
up to amplitude where $\ket{u_{x}}$ is a coherent state defined through $a_{x}\ket{u_{x}}=u_{x}\ket{u_{x}}$. We can use this classical-quantum correspondence to naturally extend the solution Eq.~(\ref{eqn_pos_1d_new}) to two- and three-dimensional systems, by constructing a position operator whose expectation value with respect to a multi-mode coherent state is reduced to the classical solution. Without loss of generality, the position operator of the radial modes of the Hamiltonian $H_{0}(t)$ in Eq.~(8) in the Appendix can be defined as  
\begin{equation}
\label{eqn_solution_single_quantum_sup}
\begin{aligned}
\textbf{\textit{x}}(t) = \mathlarger{\sum_{m = 1}^{2}} 
\frac{\textbf{\textit{b}}^{m}(t)}{Z_m} x_{0}^{m} \left(e^{i\omega_{m} t}a_{m}^{\dag} + e^{-i\omega_{m} t}a_{m}\right)
\end{aligned}
\end{equation}
where $x_{0}^{m}=\sqrt{\hbar/2M\omega_{m}}$ and $Z_m$ is a normalization constant. Note that $a_m^{\dag}$ and $a_m$ are the raising and lowering operators of an undriven time-dependent oscillator, and are not equivalent to $A_m^{\dag}$ and $A_m$ in the main text. The definition of these operators will be introduced in the following section.

\section{The forced time-dependent oscillator in the Heisenberg picture}
\label{sec_forced_sup}
The quantum mechanical position operator in the presence of an external force can be derived in two different ways. The first is based on the invariant approach~\cite{Ji_1996}, where we introduce displaced raising and lowering operators, $A^{\dag}_{m}(t)=a^{\dag}_{m}(t)-\alpha^{*}_{m}(t)$ and $A_{m}(t)=a_{m}(t)-\alpha_{m}(t)$, in terms of which Eq.~(\ref{eqn_solution_single_quantum_sup}) can be extended as
\begin{equation}
\label{eqn_solution_single_quantum_forced_sup}
\begin{aligned}
\textbf{\textit{x}}_\mathrm{q}(t) = \mathlarger{\sum_{m = 1}^{2}} 
\frac{\textbf{\textit{b}}^{m}(t)}{Z_m} x_{0}^{m} \left[A_{m}^{\dag}(t) + A_{m}(t) + 2\mathrm{Re}(\alpha_{m}(t))\right].
\end{aligned}
\end{equation}

Here, $Z_m=|\textbf{\textit{b}}^{m}(0)|$ is a normalization constant that normalizes the solution such that $|\textbf{\textit{f}}(0)|=1$. Under the approximation $|\textbf{\textit{f}}(t)|^2 \approx 1$, the raising and lowering operators are obtained as $A_{m}^{\dag}(t)= e^{i\omega_{m} t} A_{m}^{\dag}$ and $A_{m}(t)= e^{-i\omega_{m} t} A_{m}$, while the displacement $\alpha_{m}(t)$ obeys the equation
\begin{equation}
\label{eqn_displacement}
\begin{aligned}
\frac{d\alpha_{m}}{dt}+i\omega_{m}\alpha_{m}=\frac{i}{\hbar}x_{0}^{m}\textbf{\textit{F}}(t')\cdot \textbf{\textit{b}}_{0}^{m},
\end{aligned}
\end{equation}
which results in the solution
\begin{equation}
\label{alpha_def_sup}
\begin{aligned}
\alpha_{m}(t) &= e^{-i\omega_{m} t}\alpha_{m}(0) \\ &+ \frac{i}{\hbar}x_0^{m}\int_{0}^{t}dt'e^{-i\omega_{m}(t-t')}\textbf{\textit{F}}(t')\cdot \textbf{\textit{b}}_{0}^{m}
\end{aligned}
\end{equation}
with the initial condition, $\alpha_{m}(0) = x_{0}^{m} \textbf{\textit{F}}(0)\cdot \textbf{\textit{b}}_{0}^{m}/\hbar \omega_{m}$~\cite{Ji_1996}. It can be seen that the expectation value of $\textbf{\textit{x}}_\mathrm{q}(t)$ with respect to the two-mode coherent state $\ket{u_{1}u_{2}}_{A}=\ket{u_{1}}_{A}\otimes\ket{u_{2}}_{A}$ results in the classical solution $\textbf{\textit{x}}_\mathrm{c}(t)$ up to amplitude: $\bra{u_{1}u_{2}}_{A}\textbf{\textit{x}}_\mathrm{q}(t)\ket{u_{1}u_{2}}_{A} \propto \textbf{\textit{x}}_\mathrm{c}(t)$. 
Another method to obtain the forced position operator in Eq.~(\ref{eqn_solution_single_quantum_forced_sup}) is to use perturbation theory. The perturbation term $V(t)$ in Eq.~(8) in the Appendix for mode $m$ can be expressed as
\begin{equation}
\label{perturbation}
\begin{aligned}
V(t)&=-\mathlarger{\sum_{m = 1}^{2}}\textbf{\textit{F}}(t)\cdot \textbf{\textit{b}}^{m}(t) \frac{x_{0}^{m}}{Z_m} (e^{i\omega_{m}t}a_{m}^{\dag}+e^{-i\omega_{m}t}a_{m})
\\ &\approx -\mathlarger{\sum_{m = 1}^{2}}\textbf{\textit{F}}(t)\cdot \textbf{\textit{b}}^{m}_{0} \frac{x_{0}^{m}}{Z_m} (e^{i\omega_{m}t}a_{m}^{\dag}+e^{-i\omega_{m}t}a_{m})
\end{aligned}
\end{equation}
where the oscillating term $\textbf{\textit{b}}^{m}_{1} \mathrm{cos}(\omega_\mathrm{rf} t)$ has been neglected because it leads to terms proportional to $q^{2}$ in the final expression. Then, the Magnus expansion results in the propagator
\begin{equation}
\label{eqn_magnus}
\begin{gathered}
U(t)=\prod_{m=1}^{2} e^{-i\Phi_{m}(t)}D\left({\alpha}_{m}^{(1)}(t)\right), \\
{\alpha}_{m}^{(1)}(t)= \frac{i}{\hbar}x_{0}^{m}\int_{0}^{t}dt_{1} e^{i\omega_{m}t_{1}} \textbf{\textit{F}}(t_{1})\cdot \textbf{\textit{b}}_{0}^{m}\\
\begin{aligned}
\Phi_{m}(t)&=\frac{1}{\hbar^2}{x_{0}^{m}}^{2}\int_{0}^{t}dt_{1}\left(\textbf{\textit{F}}(t_{1})\cdot \textbf{\textit{b}}_{0}^{m}\right) \\ 
&\times\int_{0}^{t_1}dt_{2} \left(\textbf{\textit{F}}(t_{2})\cdot \textbf{\textit{b}}_{0}^{m}\right) \mathrm{sin}\left(\omega_{m}(t_{2} -t_{1})\right). 
\end{aligned}
\end{gathered}
\end{equation}

Acting $U(t)$ on Eq.~(\ref{eqn_solution_single_quantum_sup}), we get the forced position operator in the Heisenberg picture as
\begin{equation}
\label{eqn_solution_quantum_forced_perturbation}
\begin{gathered}
\textbf{\textit{x}}_\mathrm{q}(t) = U^{\dag}(t)\textbf{\textit{x}}(t)U(t) \\
= \mathlarger{\sum_{m = 1}^{2}} 
\frac{\textbf{\textit{b}}^{m}(t)}{Z_m} x_{0}^{m} \left[e^{i\omega_{m}t}a_{m}^{\dag} + e^{-i\omega_{m}t}a_{m} + 2\mathrm{Re}(\tilde{\alpha}_{m}(t))\right]
\end{gathered}
\end{equation}
where $\tilde{\alpha}_{m}(t)=e^{-i\omega_{m}t}{\alpha}_{m}^{(1)}(t)=\alpha_{m}(t)-e^{-i\omega_{m}t}\alpha_{m}(0)$, with $\alpha_{m}(t)$ from Eq.~(\ref{alpha_def_sup}). It can be shown that Eq.~(\ref{eqn_solution_quantum_forced_perturbation}) is equivalent to Eq.~(\ref{eqn_solution_single_quantum_forced_sup}) by substituting $a^{\dag}_{m}=A^{\dag}_{m}+\alpha^{*}_{m}(0)$ and $a_{m}=A_{m}+\alpha_{m}(0)$ into Eq.~(\ref{eqn_solution_quantum_forced_perturbation}). Therefore, we can use either $a_m$ or $A_m$ depending on the context. Note that at instance $t=0$, the oscillator states $\ket{n_{m}}_{A}$ and $\ket{n_{m}}_{a}$, where the subscripts denote the operators with respect to which the states are defined, are related by $\ket{n_{m}}_{A}=D(\alpha_{m})\ket{n_{m}}_{a}$~\cite{Ji_1996}. In particular, $\ket{0_m}_{A}=\ket{\alpha_{m}}_{a}$ since $A_{m}\ket{0_{m}}_{A}=(a_{m}-\alpha_{m})\ket{\alpha_{m}}_{a} = (\alpha_{m}-\alpha_{m})\ket{\alpha_{m}}_{a_{m}}=0$. The basis set $\{\ket{n_{m}}_{A}\}$ can actually be interpreted as a set of dressed states in the presence of the external force, \textbf{\textit{F}}(t).

\section{Light-atom interactions}
\label{sec_light-atom}
Let us consider the commonly used Hamiltonian of a trapped ion, which contains an internal two-level system and motional degree of freedom, coupled by an external radiation field. In the interaction picture of the two-level system, the Hamiltonian can be expressed as~\cite{Leibfried_2003}
\begin{equation}
\label{eqn_Hamiltonian_sup}
H_\mathrm{LA}(t)=H(t) + \frac{\hbar \Omega}{2} \left[e^{i\textbf{\textit{k}}\cdot\bm{x}(t)} e^{-i\phi(t)}\sigma^{\dag} +  \mathrm{h. c.} \right]
\end{equation}
where $H(t)$ is the Hamiltonian of the ion motion in Eq.~(8) in the Appendix, and the definition of the parameters and operators are given in the main text. If we assume that the radiation field does not drive transitions between the oscillator states, we can go into the interaction picture with respect to the entire Hamiltonian $H(t)$, through which the Hamiltonian is transformed as 
\begin{equation}
\label{eqn_Hamiltonian_sup_int}
H'_\mathrm{LA}(t)=\frac{\hbar \Omega}{2} \left[e^{i\textbf{\textit{k}}\cdot\bm{x}_\mathrm{q}(t)} e^{-i\phi(t)}\sigma^{\dag} +  \mathrm{h. c.} \right]
\end{equation}
where $\textbf{\textit{x}}_\mathrm{q}(t)$ is from either Eq.~(\ref{eqn_solution_single_quantum_forced_sup}) or Eq.~(\ref{eqn_solution_quantum_forced_perturbation}). As in the previous section, it is possible to use either $\ket{n_{m}}_{A}$ or $\ket{n_{m}}_{a}$, given the right initial conditions. On the other hand, if significant transitions between oscillator states are induced by the external field, i.e., the detuning is close to the resonant frequency of the oscillator modes, then the states $\{\ket{n_{m}}_{A}\}$ do not serve as a good basis because the perturbation due to the radiation field may become comparable to $V(t)$ in Eq.~(8) in the Appendix. In this case, we must solve the effective Hamiltonian
\begin{equation}
\label{eqn_Hamiltonian_sup_int_2}
H'_\mathrm{LA}(t)=-\textbf{\textit{F}}(t)\cdot\textbf{\textit{x}}(t) + \frac{\hbar \Omega}{2} \left[e^{i\textbf{\textit{k}}\cdot\bm{x}(t)} e^{-i\phi(t)}\sigma^{\dag} +  \mathrm{h. c.} \right]
\end{equation}
where $\textbf{\textit{x}}(t)$ is the position operator in Eq.~(\ref{eqn_solution_single_quantum_sup}), with the unperturbed basis being $\{\ket{n_{m}}_{a}\}$. Since this work is mainly focused on the carrier transition, which does not involve transitions between oscillator states, the effective Hamiltonian Eq.~(\ref{eqn_Hamiltonian_sup_int}) is a good approximation.

\begin{figure*}[ht]
\centering
\includegraphics[width=0.9\textwidth]{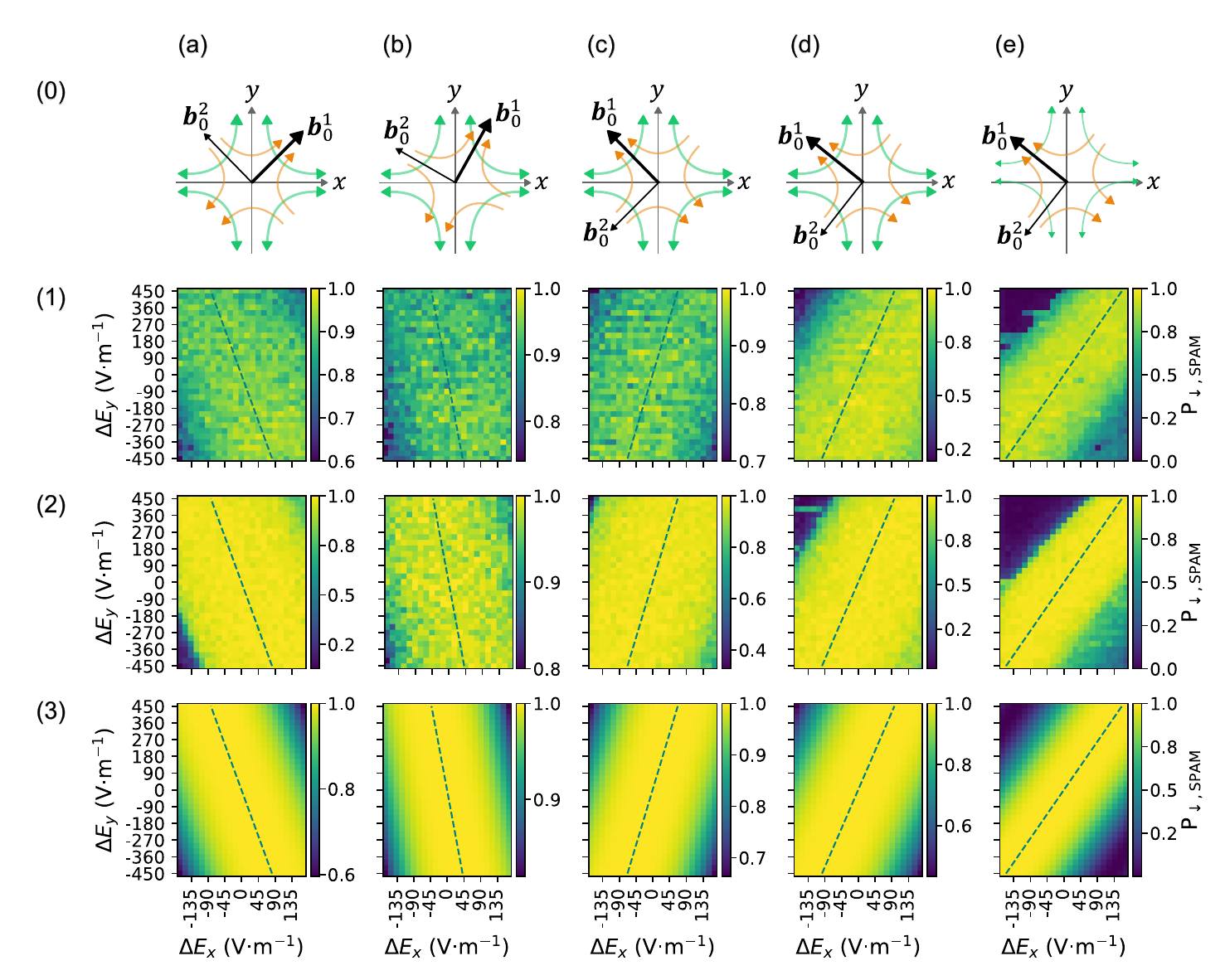}
\caption{Results of the two-dimensional displacement scan with the Doppler/SPAM beam. Columns (a)-(e) categorize results based on different potential configurations, while rows (0)-(3) present results obtained under identical conditions, differing only in the potential configuration. For each column, the secular frequencies and potential rotation angles are (a) $\{\omega_{1},\omega_{2}\}/2\pi$=\{1.79(0), 2.14(6)\} MHz, $\theta\approx\pi/4$, (b) $\{\omega_{1},\omega_{2}\}/2\pi$=\{1.87(4), 2.07(3)\} MHz, $\theta\approx\pi/3$, (c) $\{\omega_{1},\omega_{2}\}/2\pi$=\{1.82(0), 2.12(1)\} MHz, $\theta\approx3\pi/4$, (d) $\{\omega_{1},\omega_{2}\}/2\pi$=\{1.75(6), 2.22(6)\} MHz, $\theta\approx7\pi/9$, (e) $\{\omega_{1},\omega_{2}\}/2\pi$=\{1.26(8), 1.88(2)\} MHz, $\theta\approx7\pi/9$. Only the dc voltages have been varied in (a)-(d), while in (e), the rf voltage amplitude has been reduced by a factor of 0.8, keeping the dc voltages identical to those in (d). Row (0) shows the schematic depiction of potential configurations. The data in the subsequent rows are given as follows. Experimental data of the scan with the Doppler/SPAM beam detuned by (1) -22 MHz, (2) -10 MHz from the cooling transition during Doppler cooling. (3) Simulation data of the scan conducted with the Doppler/SPAM beam, $\mathrm{P}_{\downarrow,\mathrm{SPAM}}$. The theoretical rf pseudo-null lines induced by the Doppler/SPAM beam are represented as dashed lines colored in teal.}
\label{fig_2D_fluorescence}
\end{figure*}

\section{Two-dimensional displacement scan: Doppler/SPAM beam}
\label{sec_doppler}
The experimental results of the two-dimensional displacement scan corresponding to the simulation results in row (4) of Fig. 2 of the main text is presented in Fig.~\ref{fig_2D_fluorescence}. Rows (0) and (3) are identical to rows (0) and (4) of Fig. 2 in the main text, shown for convenience. In row (0), the Doppler/SPAM beam is detuned by -22 MHz, the rf frequency, from the cooling transition during Doppler cooling, while in row (1), it is detuned by -10 MHz, the optimal Doppler cooling condition.

\begin{figure}[ht]
\centering
\includegraphics[width=1\columnwidth]{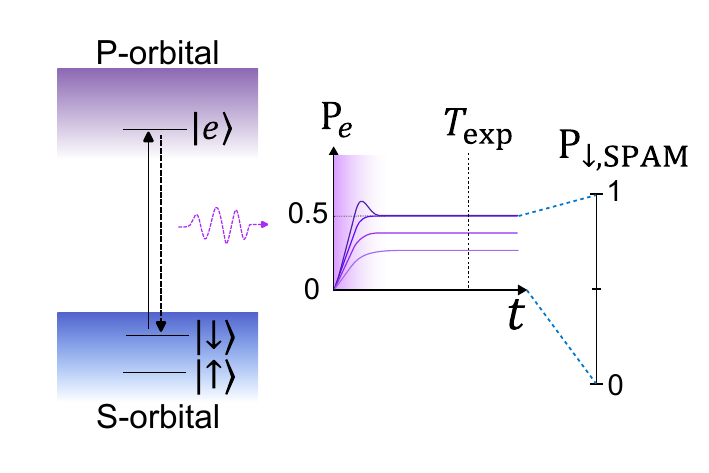}
\caption{Schematic depiction of the derivation of $\mathrm{P}_{\downarrow,\mathrm{SPAM}}$. On the left, the internal states $\ket{\uparrow}$, $\ket{\downarrow}$, and the state $\ket{e}$, used for the dissipative scattering processes, are visualized in the S- and P-orbitals, respectively. The Doppler/SPAM beam drives transitions between $\ket{\downarrow}$ and $\ket{e}$. The population of $\ket{e}$, $\mathrm{P}_{e}(t)$, saturates at a level set by the Rabi frequency modified by micromotion, $\tilde{\Omega}$, and the decay rate of $\ket{e}$, given by $\Gamma$. Its maximum value is 1/2, and decreases as $\tilde{\Omega}$ is reduced, as shown at the center of the figure. The steady state value $\mathrm{P}_{e}(t\rightarrow\infty)$ is translated into the probability used for state detection, $\mathrm{P}_{\downarrow,\mathrm{SPAM}}$, with the scaling factor directly determined by the Rabi frequency modified by micromotion.}
\label{fig_P_scat}
\end{figure}

Here, we elaborate on the measurement of $\mathrm{P}_{\downarrow,\mathrm{SPAM}}$, which can be extracted by conducting state detection immediately after Doppler cooling. Experimentally, it is the probability to correctly detect the fluorescing ion as the state $\ket{\downarrow}$. To comprehend the full picture including photon scattering, consider the schematic energy level diagram shown in Fig.~\ref{fig_P_scat}~\cite{Olmschenk_2007}, with the states $\ket{\uparrow}$ and $\ket{\downarrow}$ in the S-orbital, and $\ket{e}$ in the P-orbital. In realistic conditions, various sidebands are present and different states in the S- and P-orbitals are involved in each step of state preparation and measurement (SPAM), but the simplistic depiction in Fig.~\ref{fig_P_scat} is sufficient for our purposes. 

Let us assume a two-level system composed of the states $\ket{\downarrow}$ and $\ket{e}$, where the decay rate of $\ket{e}$ is given as $\Gamma$. When light drives transitions between the two states at a rate given by $\Omega$, the steady state population of the state $\ket{e}$ is derived as~\cite{DeVoe_1989}
\begin{equation}
\label{eqn_P_e}
\mathrm{P}_{e}(t\rightarrow\infty)=\frac{1}{2}\frac{s}{1+s}, \ \ 
s\approx\frac{2(\tilde{\Omega}/\Gamma)^2}{1+(2\Delta/\Gamma)^2},
\end{equation}
where $\tilde{\Omega}=J_{0}(\beta)\Omega$ denotes the modified Rabi frequency, parameterized by the Bessel function of type zero and the modulation depth $\beta$, in the presence of micromotion at a given displacement. Additionally, we assume $\Delta=0$. In the typical regime of operation where $\beta=0$ and $\Omega\gg\Gamma$, we have $\mathrm{P}_{e}(t\rightarrow\infty)\approx1/2$ in the steady state. On the other hand, when the net phase modulation from micromotion is significant, $\beta\gg1$ and $\tilde{\Omega}\ll \Gamma$, we get, $\mathrm{P}_{e}(t\rightarrow\infty)\propto\ |\tilde{\Omega}|^{2}$, varying with the value of $\beta$. This is graphically visualized in the plot of $\mathrm{P}_{e}(t)$ at the center of Fig.~\ref{fig_P_scat}. The transient response of $P_{e}(t)$, represented as the shaded area in light purple, is generally inaccessible during experiments due to the large decay rate $\Gamma/2\pi\approx$ 20 MHz. During detection, the ion's fluorescence is collected for a fixed exposure time, indicated as $T_\mathrm{exp}$ in Fig.~\ref{fig_P_scat}, and the resulting Poisson distribution of photon counts is used for state discrimination between $\ket{\uparrow}$ and $\ket{\downarrow}$~\cite{Olmschenk_2007}. Recall that $\epsilon=1-\mathrm{P}_{\downarrow,\mathrm{SPAM}}$ is not equivalent to the probability of detecting $\ket{\uparrow}$, as the ion is always fluorescing by transitioning between $\ket{\downarrow}$ and $\ket{e}$ in the steady state. However, since the two cases are experimentally indistinguishable, the standard state detection scheme can measure $\mathrm{P}_{\downarrow,\mathrm{SPAM}}$ from the observed intensity variations of fluorescence.

Clearly, the patterns of $\mathrm{P}_{\downarrow,\mathrm{SPAM}}$, obtained from both the system Hamiltonian simulation and the more realistic model incorporating scattering, are predominantly governed by the same displacement-dependent variation in Rabi frequencies. In the latter, which relates more directly to the observed quantity, $\mathrm{P}_{\downarrow,\mathrm{SPAM}}$ may be seen as a map from the domain $(0,0.5)$ of $\mathrm{P}_{e}$ to the codomain $(0,1.0)$, scaled by the micromotion-modified Rabi frequency, as shown in Fig.~\ref{fig_P_scat}. Our claim is supported by the excellent agreement of the experimental data of $\mathrm{P}_{\downarrow,\mathrm{SPAM}}$ in rows (1) and (2) of Fig.~\ref{fig_2D_fluorescence} with the simulation results in row (3). When the ion is located along the rf pseudo-null line induced by the Doppler/SPAM beam, the fluorescence is largest, i.e., $\mathrm{P}_{\downarrow,\mathrm{SPAM}}=1$, leading to efficient Doppler cooling and higher detection fidelity. However, when it is displaced from the pseudo-null and $\phi_\mathrm{mm}(t)\neq0$, the value of $\mathrm{P}_{\downarrow,\mathrm{SPAM}}$ decreases, preventing the ion from reaching the Doppler limit~\cite{Berkeland_1998} and resulting in larger measurement errors.

The rf pseudo-null induced by the Doppler/SPAM beam can clearly be seen throughout Fig.~\ref{fig_2D_fluorescence} (a)-(e), with the relative strengths between different displacement values in good agreement with simulation results. There are, however, regions where $P_{\downarrow,\mathrm{SPAM}}$ drops dramatically to near zero in all cases. We observe a significant increase in the overall ion motion in these regions, detectable via the electron-multiplying charge-coupled device (EMCCD) used for ion imaging. Our conjecture is that due to inefficient Doppler cooling in these areas, the ion image becomes blurred, resulting in a significant loss in fluorescence. These effects are noticeably reduced when the Doppler/SPAM beam is tuned to the rf frequency, as shown in row (1), rather than under optimal Doppler cooling conditions in row (2), consistently explaining the results in Fig. 2 of the main text. Note that the detection fidelity is slightly lower in row (1) compared to row (2) due to constraints in the optical devices, but this does not compromise the validity of the experimental data.

\begin{table*}[t]
    \centering
    \renewcommand{\arraystretch}{1.2} 
    \setlength{\tabcolsep}{10pt} 
    \begin{tabular}{c c c c c c}
        \toprule\toprule
         & (a) & (b) & (c) & (d) & (e) \\ 
        \midrule
        $\alpha \ (\mathrm{cm}^{-2})$  & -980  & -550  & -850  & \multicolumn{2}{c}{-1370}  \\
        \midrule
        $\beta, \ \gamma \ (\mathrm{cm}^{-2})$  & \multicolumn{5}{c}{134, 2200} \\
        \midrule
        $V_{\mathrm{dc}} \ (\mathrm{V})$  & \multicolumn{5}{c}{5} \\
        \midrule
        $V_{\mathrm{rf}} \ (\mathrm{V})$  & \multicolumn{4}{c}{200} & 160  \\ 
        \midrule
        $\theta \ (\mathrm{radians})$  & $\pi/4$  & $\pi/3$  & $3\pi/4$  & $7\pi/9$  & $7\pi/9$  \\ 
        \midrule
        $\omega_{1}/2\pi \ (\mathrm{MHz})$  & 1.79(0)  & 1.87(4)  & 1.82(0)  & 1.75(6)  & 1.26(8)  \\
        $\textbf{\textit{b}}_{0}^{1} \ (-)$  & (0.707, 0.707)  & (0.497, 0.868)  & (-0.707, 0.707)  & \multicolumn{2}{c}{(-0.767, 0.642)}  \\
        $\textbf{\textit{b}}_{1}^{1} \ (-)$  & (0.045, -0.045)  & (0.0316, -0.0551)  & (-0.045, -0.045)  & (-0.0496, -0.0416)  & (-0.0397, -0.0333)  \\
        \midrule
        $\omega_{2}/2\pi \ (\mathrm{MHz})$  & 2.14(6)  & 2.07(3)  & 2.12(1)  & 2.22(6)  & 1.88(2)  \\
        $\textbf{\textit{b}}_{0}^{2} \ (-)$  & (0.707, -0.707)  & (0.868, -0.497)  & (-0.707, -0.707)  & \multicolumn{2}{c}{(0.642, 0.767)}  \\
        $\textbf{\textit{b}}_{1}^{2} \ (-)$  & (0.045, 0.045)  & (0.0551, 0.0316)  & (-0.045, 0.045)  & (0.0416, -0.0496)  & (0.0333, -0.0397)  \\
        \midrule
        $x_{\mathrm{max}}$ (\textmu m)  & $\pm$307  & $\pm$279  & $\pm$304  & $\pm$315  & $\pm$556  \\
        $y_{\mathrm{max}}$ (\textmu m)  & $\pm$838  & $\pm$898  & $\pm$830  & $\pm$798  & $\pm$1336  \\
        \bottomrule\bottomrule
    \end{tabular}
    \caption{Numerical values of the parameters used for the potential configurations in the simulation of the two-dimensional displacement scan}
    \label{table_params}
\end{table*}

\section{Simulation parameters}
\label{sec_sim_params}
Numerical values of the parameters used for the simulations in Fig. 2 of the main text and Fig.~\ref{fig_2D_fluorescence} are listed in Table.~\ref{table_params}. The columns (a)-(e) label different potential configurations. The secular-micromotion modes $\{\omega_{m}, \textbf{\textit{b}}_{0}^{m}$, $\textbf{\textit{b}}_{1}^{m}\}$ are obtained from the parameters listed in the top five rows. The maximum displacement of the ion, $x_{\mathrm{max}}$, $y_{\mathrm{max}}$, induced by the control electric fields, $\Delta E_{x}, \ \Delta E_{y}$ are given in the last row in terms of spatial units. They are well within the spatial extent of the beam width and Rayleigh range of the lasers used for experimentation. Note, calculations have been conducted in cgs units. The potential rotation $\theta$ and magnitude of the control electric fields, $\Delta E_{x}, \ \Delta E_{y}$, have been extracted from COMSOL, whereas the secular frequencies $\omega_{x}, \  \omega_{y}$ have been measured through sideband scans via the Raman transition~\cite{Leibfried_2003}.

\bibliography{ref}

\end{document}